# A geometrical aperture-width relationship for rock fractures

Behzad Ghanbarian[1*], Edmund Perfect[2] and Hui-Hai Liu[3]

[1] Department of Geology, Kansas State University, Manhattan 66506 KS, USA

[2] Department of Earth & Planetary Sci., University of Tennessee, Knoxville 37996 TN, USA

[3] Aramco Research Center, 16300 Park Row, Houston 77084 TX, USA

[*] Corresponding author's Email address: ghanbarian@ksu.edu

**Abstract**

The relationship between fracture aperture (maximum opening; $d_{max}$) and fracture width ($w$) has been the subject of debate over the past several decades. An empirical power law has been commonly applied to relate these two parameters. Its exponent ($n$) is generally determined by fitting the power-law function to experimental observations measured at various scales. Invoking concepts from fractal geometry we theoretically show, as a first-order approximation, that the fracture aperture should be a linear function of its width, meaning that $n = 1$. This finding is in agreement with the result of linear elastic fracture mechanics (LEFM) theory. We compare the model predictions with experimental observations available in the literature. This comparison generally supports a linear relationship between fracture aperture and fracture width, although there exists considerable scatter in the data. We also discuss the limitations of the proposed model, and its potential application to the prediction of flow and transport in fractures. Based on more than 170 experimental observations from the literature, we show that such a linear relationship, in combination with the cubic law, is able to scale flow rate with fracture aperture over ~14 orders of magnitude for variations in flow rate and ~5 orders of magnitude for variations in fracture width.

1. Introduction

Modeling flow and transport within fracture networks requires knowledge of fracture attributes and their scaling properties. A fracture is typically characterized by its aperture, width, displacement, and surface roughness. It is well documented in the literature that the surfaces of natural rock fractures are rough and follow self-affine scaling from fractal geometry[1–4]. The self-affinity of the surfaces of fractures is best understood in a statistical sense[5]. According to self-affinity, if one considers the height difference between two points separated by distance $|x_i - x_j|$ on a self-affine surface, then

$$\langle h(x_i) - h(x_j) \rangle \propto |x_i - x_j|^H \tag{1}$$

where $h(x_i)$ and $h(x_j)$ are the heights at points $x_i$ and $x_j$, respectively, and $H$ is the Hurst exponent characterizing the surface roughness. Generally speaking, the larger the Hurst exponent, the smoother the fracture surface. A value of $H \sim 0.8$ was reported for granite[6], while $H \sim 0.5$ for sandstones[7]. However, a wider range of the Hurst exponent e.g., $0 < H < 0.9$ has been reported for a variety of rock joints[5].

One of the most important characteristics of an individual fracture is its aperture, defined herein as the maximum opening ($d_{max}$). The maximum opening is related to the average opening via $d_{max} = 4/\pi \, d_{ave}$[8]. Although aperture has been typically characterized as either the maximum or the average opening, due to the rough surface of fractures, there exists a distribution of aperture openings, rather than a single unique value for a given fracture. It is worth mentioning that the aperture opening distribution can only be inferred once the fracture surfaces have been characterized, which requires accurate imaging of the surface roughness.

Fracture aperture has been shown to control fluid flow and solute transport processes. For example, the well-known cubic law[9,10] relates flow in a single fracture to the product of fracture width $w$ and fracture aperture raised to the power three (i.e., $Q \propto w d_{ave}^3 \nabla h \propto w d_{max}^3 \nabla h$ in which $Q$ is flow rate and $\nabla h$ is hydraulic gradient). Accordingly, the scaling relationship between fracture aperture and its width has been the subject of active research and debate over the past several decades. We should point out that the terms width and length have been interchangeably used in the literature to describe the straight line from one tip of a rock fracture to another, which can be a source of confusion.

A commonly applied model linking fracture aperture to fracture width is the following empirical power law[11,12]:

$$d_{max} = cw^n \tag{2}$$

in which $c$ is a constant coefficient, $d_{max}$ is the fracture aperture as previously defined, $w$ is the fracture width, and $n$ is an empirical exponent. Different values of $n$ have been reported in the literature based on experimental observations, stochastic reasoning, and theoretical models. In Table 1, we summarize the various values of *n* (and if appropriate the corresponding coefficients of determination, $R^2$, determined by directly fitting Eq. (2) to measurements), that have been obtained using these three different approaches. For a comprehensive review, see Bonnet et al.[11].

From a theoretical perspective, linear elastic fracture mechanics (LEFM) theory, which has been successfully applied to rock fractures, predicts that the fracture aperture should scale linearly with fracture width[13], meaning that the value of *n* in Eq. (2) should be equal to 1. A relationship similar to Eq. (2) was suggested by Oron and Berkowitz[14] between mean aperture and mean width for varying length sections within a self-affine fractal fracture. According to Oron and Berkowitz[14], *n* = *H*, the Hurst exponent that characterizes the roughness of the fracture surface (see their Eq. 24).

It is possible to stochastically relate fracture aperture to fracture width[15]. Based on this approach, it is well documented in the literature (see e.g., Bonnet et al.[11]) that fracture width- and aperture-size distributions both conform to power-law probability density functions (i.e., $f(d_{max}) = C_d d_{max}^{-\alpha}$ and $f(w) = C_l w^{-\beta}$ in which $C_d$ and $C_l$ are constant coefficients). Assuming that $d_{max}$ and *w* are related via the mathematical power-law function, Eq. (2), one can write $f(d_{max})d(d_{max}) = f(w)dw$. Given that $d(d_{max})/dw = cnw^{n-1}$ from Eq. (2), then $n = (\beta - 1)/(\alpha - 1)$. Scholz and Cowie[15] reported $\alpha = 2.2$ and $\beta = 2.1$ and found $n \approx 1$ for a variety of tectonic environments. Those authors also showed that *n* = 1 could accurately represent their measured data. Besides power law scaling[16,17], the log-normal probability density function has also been used to describe distributions of fracture aperture and/or width[18,19].

In addition to theoretical and stochastic values for $n$, there are numerous empirical estimates of $n$ in the literature. For example, Hatton et al.[20] experimentally analyzed two datasets (i.e., Kelduhverfi and Myvatn) from the Krafla fissure swarm, Iceland, and found a break in slope in the aperture-width data when plotted on a log-log scale. For the Kelduhverfi dataset, they reported $n = 2.2 \pm 0.24$ with $R^2 = 0.41$ for $w < 3$ m and $n = 0.89 \pm 0.08$ with $R^2 = 0.76$ for $w > 3$ m, while for the Myvatn dataset, $n = 1.78 \pm 0.47$ with $R^2 = 0.22$ for $w < 3$m and $n = 0.63 \pm 0.08$ with $R^2 = 0.81$ for $w > 3$m. The break in slope was interpreted as an indication of scale-dependent growth mechanisms. In contrast, Renshaw and Park[21] argued that, "… the break in slope is instead intrinsic to the fracturing processes and represents the maximum length scale at which the apertures of smaller fractures are affected by stress perturbations induced by larger fractures." Later, Main et al.[22] reanalyzed the Kelduhverfi dataset using a new method based on Schwartz's Information Criterion and a Bayesian approach and found a break in slope near 12 m (instead of 3 m reported by Hatton et al.[20]) as well as $n = 1.49 \pm 0.29$ for $w < 12$ m and $n = 0.64 \pm 0.41$ for $w > 12$ m. In Main et al.[22] the difference in the Bayesian information criterion (BIC) between the double- and single-slope models, conditional on the optimal change point, was about 5 (see their Fig. 1). This implies a relative likelihood of $\exp(5/2)$, meaning that the double-slope model is nearly 12 times more likely than the single-slope model to be correct for the Hatton et al.[20] data set (Ian Main, 2018; personal communication).

Other values of $n$, estimated by fitting data sets obtained over a variety of length scales, are listed in Table 1. Interestingly, the arithmetic average of all of the experimentally-determined $n$ values in Table 1 is $1.05 \pm 0.07$, a value which is not statistically different from unity. It should be noted that most *n* values presented in Table 1 are based on rock fracture measurements. However, $n = 0.47 \pm 0.03$ was derived by Walmann et al.[23] from cracks in clayey soils, which are unconsolidated as compared to rocks.

The main objective of this study is to use concepts from fractal geometry to develop a first-order approximation of the relationship between fracture aperture $d_{max}$ and fracture width *w*. To our knowledge, no such approach has previously been proposed to predict the $d_{max}$-*w* relationship. In what follows, we briefly describe fractal geometry, introduced by Benoit B. Mandelbrot[24], and present its fundamental concepts. We then derive a geometrical relationship between the aperture and width of a single rough-walled fracture, resulting in a geometrically-based prediction of the exponent *n* in Eq. (2). Finally, we compare our theoretical results with experimental observations reported in the literature and relate them to flow and transport models.

**2. Theory**

Fractal geometry, introduced by Benoit B. Mandelbrot[24], has been shown to be a robust and appropriate approach for modeling the multiscale structure of complex and heterogeneous media such as rocks, soils, and fracture networks[5,25–32]. A fractal object is characterized by having a (typically non-integer) dimension less than the Euclidean dimension of the space it is embedded in. If a fractal object is rescaled in all directions with the same scaling factor, a statistically similar object is reproduced; this property is termed self-similarity.

Many natural objects are self-affine rather than self-similar, meaning that they require different scaling factors in different directions[5]. The concept of self-affinity has been used to model the physical and geometrical properties of fracture networks[4,14,33]. It has also been applied to statistically characterize fracture surfaces[34]. Poon et al.[35], among others (see Sahimi[5], for a comprehensive review), modeled surface roughness of fractures by means of self-affinity. However, none of these models predict the relationship between fracture aperture $d_{max}$ and fracture width $w$.

In the simple model that we present, self-affinity is assumed. Let us presume that the cross-section of a natural fracture can be represented by an ellipse with a rough boundary, as shown in Fig. 1. It should be pointed out that an unembroidered ellipse is the shape of a Griffith crack traditionally employed in linear elastic fracture mechanics (see e.g., Zimmerman and Main[36]). Other crack geometries e.g., edge, corner, or semi-circular have also been used within Griffith theory. We further assume that the rough boundary is fractal, and thus, following Mandelbrot[25], one may relate the fracture perimeter $P$ to its area $A$ as follows:

$$P \propto A^{\frac{D_b}{2}} \qquad (3)$$

where $D_b$ is the boundary fractal dimension ($1 \leq D_b < 2$) characterizing the roughness of a cross-section taken through a fractal fracture surface (Fig. 1). The higher the $D_b$ value, the rougher the boundary.

Equation (3) has been successfully applied to relate perimeter to area in clouds [37], metallic rough fractures[38], and soil and rock pores[39,40]. Furthermore, it has been also used in combination with the model of Patzek and Silin[41] to describe fluid flow in tubular pores with rough surfaces, and to accurately estimate water relative permeability[42].

We can compute the width of the fractal perimeter of a rough-boundary ellipse shown in Fig. 1, by invoking the Mandelbrot[25] approach, followed by Wheatcraft and Tyler[43] and many others. Based on this approach, a fractal length $L_f$ is a function of some measurement scale $\varepsilon$ and the straight-line distance $L_s$ between the two ends of the fractal path as follows:

$$L_f = \varepsilon^{1-D_l} L_s^{D_l} \tag{4}$$

in which $D_l$ is the fractal dimension of the fractal length. Equation (4) is valid for both self-similar and self-affine fractal curves. For a self-similar fractal curve, divider method, box counting, and mass scaling estimates of $D_l$ will all be the same. However, for a self-affine fractal curve, different values of $D_l$ will be obtained using different evaluation methods[44]. We should also note that the concept underlying fractals is self-similarity or self-affinity, that is, invariance against variations in scale or size (scale-invariance). Accordingly, the fractal dimension is theoretically scale-invariant. However, in the nature, objects are only approximately fractal and scale invariant, and one may expect the fractal dimension to vary from one scale to another. A notable example is fractal dimension determination from images. For example, Baveye et al.[45] demonstrated the effects of image resolution, thresholding, and algorithm used to generate binary images on the estimation of fractal dimension.

Assuming that the rough boundary shown in Fig. 1 is a fractal length, setting $D_l = D_b$ and $L_s = w$ in Eq. (4) in combination with $P \propto L_f$ yields:

$$P \propto \varepsilon^{1-D_b} w^{D_b} \tag{5}$$

Equation (5) necessarily means that perimeter is measurement scale-dependent. Given that $1 \leq D_b < 2$, $P$ is directly proportional to the fracture width $w$, while inversely proportional to the measurement scale $\varepsilon$. As a consequence, when $\varepsilon$ tends to zero, $P$ approaches infinity.

The area of an ellipse with a smooth boundary (represented by the black dashed line in Fig. 1) is a function of the product of its semi-minor and semi-major axes ($wd_{max}$). Accordingly, for the rough-boundary ellipse (shown in red) one can approximately set

$$A \propto wd_{max} \tag{6}$$

Note that Walmann et al.[23] also used a relationship similar to Eq. (6) to relate the area of a fracture to its aperture and width.

Substituting Eqs. (5) and (6) into Eq. (3) gives

$$d_{max} = Cw \tag{7}$$

where $C$ is a numerical prefactor whose value depends on the measurement scale $\varepsilon$ (i.e., $C \propto \varepsilon^{\frac{2-2D_b}{D_b}}$).

Equation (7), represents a geometrical scaling relationship linearly relating fracture aperture to fracture width. It is in agreement with the linear elastic fracture mechanics (LEFM) approach, which predicts that the fracture aperture should scale linearly with fracture width[13]. Within the LEFM theory framework, the coefficient $C$ would be equal to $\sigma(1 - \nu)/\mu$ in which $\sigma$ is the effective driving stress (remote tension plus the internal fluid pressure), $\nu$ is Poisson's ratio, and $\mu$ is the shear modulus[13].

As the derivation of Eq. (7) shows the roughness exponent cancels meaning that it is mainly the fracture geometry governing the linear $d_{max}$-$w$ relationship rather than the roughness per se. It is interesting that the rough-boundary ellipse produces the same result as the smooth-boundary ellipse applied by LEFM. Although our geometrical terminology is different from that of LEFM, the obtained results indicate that perturbing the boundary of a smooth-boundary ellipse, used in LEFM theory, with self-affinity duplicates the linear $w$-$d_{max}$ relationship. In Appendix A, using the same geometrical terminology, we, however, demonstrate that there exists no simple linear relationship between fracture aperture and its width for a smooth-boundary ellipse, unless $w$ is significantly greater than $d_{max}$. In Appendix B, we assume a rectangular fracture embroidered with a fractal rough boundary (see Fig. B1) and demonstrate that when $w \gg d_{ave}$ the average fracture aperture ($d_{ave}$) should scale linearly with its width ($w$).

Equation (7) is also in agreement with the stochastic approach of Scholz and Cowie[15] and the average of the experimentally-determined $n$ values presented in Table 1. However, there are some exponents reported in the literature (e.g., those from Hatton et al.[20]) that apparently do not match the theoretical prediction of $n = 1$. In what follows, we compare the geometrical model with two datasets from Hatton et al.[20] and demonstrate that Eq. (7), with different numerical prefactor $C$ values, is able to accurately match the experimental measurements by Hatton et al.[20].

### 3. Comparison with Hatton et al. experiments

Hatton et al.[20] collected two datasets, namely Kelduhverfi and Myvatn, from the Krafla fissure swarm, one of five volcanotectonic systems in the active rift zone of northeast Iceland. It is probably worth pointing out that the Kelduhverfi and Myvatn exposures have heavy joint sets at a characteristic scale, that control the fracturing (see Fig. 2 of Hatton et al.[20]). However, this is not always the case (Ian Main, 2018; personal communication). Hatton et al.[20] measured fracture width, the straight line from tip to tip, and fracture aperture, the maximum opening displacement along the width of a fracture, using a tape measure. A total of 72 and 42 fractures were digitized for the Kelduhverfi and Myvatn areas, respectively, from Fig. 3 in Hatton et al.[20]. The yield of 72 for Kelduhverfi is less than the 79 stated in Hatton et al.[20]. The difference between the two might be due to low figure quality or duplications in the measurements.

Fracture aperture as a function of fracture width is shown on a log-log scale for the two areas in Fig. 2. As can be observed in Fig. 2a, the data points are highly scattered, not only at short length scales but also over the entire range of length scales. For example, at short length scales a fracture aperture of 0.001m corresponds to a wide range of fracture widths from near 0.1 to 1m, almost one order of magnitude. Such scatter in the data causes substantial uncertainties in the optimization of any mathematical function's parameters through a direct fitting process.

In the Kelduhverfi dataset (Fig. 2a) both fracture aperture and width span nearly four orders of magnitude. However, in the Myvatn dataset, fracture width spans nearly three and half orders of magnitude while aperture only varies over approximately two orders of magnitude. Nonetheless, as we show in Fig. 2, Eq. (7) with various numerical prefactors can accurately capture the trends in the observations. We plotted Eq. (6) with $C = 0.001$, 0.01, and 0.1 in Fig. 2; however, one can clearly see that other values of $C$ might be relevant to some data points. In both plots shown in Fig. 2, the envelope of estimated fracture aperture values via $C = 0.001$, 0.01, and 0.1 coincides remarkably with results from the two datasets from Hatton et al.[20].

Our results presented in Fig. 2 are in agreement with those of Schultz et al.[46] who experimentally demonstrated for 14 datasets that fracture aperture should linearly scale with fracture width (see their Fig. 1), similar to Eq. (6). In their Fig. 2, however, they discuss that for some experiments the value $n = 0.5$ might be more relevant than $n = 1$. Interestingly, Schultz et al.[46] found that the same numerical prefactors $C = 0.001$, $0.01$, and $0.1$ could accurately capture aperture-width trends for more than 170 fractures from 14 datasets including normal faults, strike-slip faults, and thrust faults.

**4. Comparison with flow rate experiments**

The correlation between aperture and width has an impact on flow rate $Q$ in a single fracture, which can be approximated by the following cubic law[9,10]

$$Q \propto w d_{ave}^3 \nabla h \propto w d_{max}^3 \nabla h \tag{8}$$

For the validity of the cubic law in smooth and rough fractures, see Sahimi[5] (p. 156 and 416-417) for a review as well as Neuman[47]. Equation (8) clearly shows that flow rate $Q$ is mainly controlled by fracture aperture $d_{max}$ rather than fracture width $w$. This is because in Eq. (8) aperture is raised to the power three, while width is only raised to the power of unity. However, the effect of fracture width might not be negligible, as we discuss in the following paragraphs.

If the fracture aperture and the fracture width are uncorrelated, Eq. (8) in its present form ($Q \propto d_{max}^3$) should accurately scale flow rate with aperture in a single fracture. However, if $d_{max}$ and $w$ are correlated via Eq. (2), then $w$ in Eq. (8) can be replaced with $d_{max}^{\frac{1}{n}}$, and accordingly Eq. (8) changes to

$$Q \propto d_{max}^{3+\frac{1}{n}}, \tag{9}$$

and the values $n = 0.5$[8,46,48] and $n = 1$ (see Table 1 and Eq. 7) result in $Q \propto d_{max}^5$ and $Q \propto d_{max}^4$, respectively. Equation (9) clearly shows deviation from the cubic law that should be attributed to the aperture-width relationship (Eq. 2). However, as we discuss later, there is evidence in the literature indicating that such a deviation can be attributed to surface roughness. Accordingly, Eq. (9) will likely only provide accurate results when the aperture is wide enough so that the effect of surface roughness is minimized. In what follows, we compare $Q \propto d_{max}^4$ with data from flow rate experiments.

Klimczak et al.[48] collected 8 datasets from various studies in the literature including Shiprock dikes, Florence Lake veins, Culpeper veins, Moros joints, Lodeve veins, Donner Lake dikes, Emerald Bay veins, and Emerald Bay dikes (see their Fig. 6). Measured flow rate (m³/s) versus measured fracture aperture (m) are shown in Fig. 3 on a log-log scale. As can be observed, flow rate and aperture span near 16 and 6 orders of magnitude on the vertical and horizontal axes, respectively. Generally speaking, the measurements can be grouped into two classes: (1) Shiprock dikes (represented by triangles in Fig. 3), and (2) all others. Although both classes have similar slopes on a log-log scale, the Shiprock dikes require a larger aperture than other types of fractures shown in Fig. 3 to return the same flow rate value.

In Fig. 3, we also show Eq. (9) with *n* = 1 (from Eq. 7) and two different numerical prefactors i.e., $10^2$ and $10^8$. As can be seen, $Q = 10^2 d_{max}^4$ and $Q = 10^8 d_{max}^4$ precisely scale the measurements over 14 orders of magnitude variations in flow rate and 5 orders of magnitude of fracture widths. This result is a definite improvement on the $Q \propto d_{max}^5$ scaling propounded by Klimczak et al.[48] (see their Fig. 6).

It is worth pointing out that one should expect Eq. (9) to be valid only under the laminar flow (low Reynolds number) condition. However, some of the high discharge experiments of Klimczak et al.[48] presented in Fig. 3 would likely violate this condition for water. Those experiments apparently correspond to magma flow through dikes with wide apertures. Given that magma's viscosity is remarkably high and significantly greater than that of water, the Reynolds numbers associated with those experiments should not be too large. However, magmas are typically quasi- or non-Newtonian, for which applying Darcy's law causes uncertainties in calculations. Given the lack of detailed information on the flow rate experiments of Klimczak et al.[48], care should be taken in interpreting the close correspondence between Eq. (9) and the data in Fig. 3 as model validation. Clearly, additional experimental work on water flow rates in fractures is needed in order to fully evaluate our proposed model.

Although we have shown that a simple geometrical scaling relationship, Eq. (7), in combination with the cubic law can accurately represent the experimental data from Klimczak et al.[48], there is experimental and numerical evidence in the literature that surface roughness can remarkably affect fluid flow[49]. Accordingly, an exponent different than 4 might better scale flow rate $Q$ with fracture aperture $d_{max}$. For example, lattice-gas simulations in single self-affine (anisotropic) fractures by Zhang et al.[34] revealed exponents greater than 4. For three-dimensional fractures with relatively rough ($H = 0.8$) and rough ($H = 0.3$) surfaces, those authors found exponents 2.67 and 3.15, respectively. With further decreases in $H$ they found significant increases in the exponent scaled permeability with mean aperture. Zhang et al.[34] stated that, "Our study shows *quantitatively* that the experimentally observed deviation from the cubic law can be attributed to surface roughness." Sahimi[5] also stated that, in fractures with rough surfaces, an exponent as large as 6 might be expected. Eker and Akin[50] simulated fluid flow through two-dimensional fractures with rough surfaces via the lattice-Boltzmann technique. Those authors reported that as the mean aperture-fractal dimension ratio increased, permeability increased. They stated that, "The resulting permeability values were less than the ones obtained with the cubic law estimates." Eker and Akin[50] also found that the exponent scaled permeability with mean aperture was linearly correlated to fractal dimension characterized fracture surface roughness for isotropic fractures (see their Eq. 12). More specifically, their simulations resulted in exponents ranging from 4.27 to 5.66, in accord with the results of Zhang et al.[34]. Recently, Liu et al.[51] combined concepts from fractal curves, Eq. 4, with the cubic law, Eq. 8, and proposed $Q \propto d_{ave}^{6-D_T}$ in which $D_T$ represents the tortuosity fractal dimension. Given that theoretically $1 \leq D_T < 3$, the exponent $6 - D_T$ can be expected to range between 3 and 5, with $D_T = 2$ corresponding to our model prediction. Further research on comparing these different approaches could be valuable.

## 5. Discussion

One of the main assumptions in our theoretical derivation is that the boundary of the fracture surface is fractal. Many studies in the literature indicate that fracture surfaces are rough and obey fractal geometry[1,2,52–55]. However, not every rock fracture surface is necessarily fractal at all scales. In fact, natural fractures that exhibit self-similarity or self-affinity might lose their fractal properties above and below upper and lower cutoff scales. The aperture-width relationship for fractures whose surface is non-fractal may well deviate from the theoretical linear function given in Eq. (7), as we show in Appendix A.

In addition, we assumed that fractures are elliptical with a rough boundary. Although natural fractures have been frequently represented by ellipses [56–59], in reality they typically have irregular cross sections with converging-diverging opening geometry. Furthermore, rather than a unique aperture for a given fracture, there exists a distribution of gap sizes due to the rough surfaces of fractures. Thus, we should caution that any significant deviation from elliptical shape may cause uncertainties in our proposed scaling relationship.

Madadi et al.[60] studied fluid flow in two-dimensional fractures with rough self-affine surfaces, quantified with the Hurst exponent $H$, using the lattice-Boltzmann method. They used five values of $H$ ranging from less than 0.5 (which generated rough surfaces) to near 1 (which provided relatively smooth surfaces). In their study, the mean aperture of the fractures was varied to generate narrow to wide fractures. They stated that, "Using a simple mean aperture for representing a fracture with rough internal surfaces will always result in gross errors, unless, of course, the fractures are wide, in which case the surface roughness does not really matter."

A rough profile (i.e., the fractal path shown in Fig. 1) can be characterized by two parameters: (1) the boundary fractal dimension (or Hurst exponent), and (2) the root-mean-square of the roughness height. The former characterizes the boundary roughness, while the latter controls the roughness thickness. The higher the boundary fractal dimension, the rougher the fractal path. Likewise, the higher the root-mean-square of the roughness height, the thicker the fractal profile. In our theoretical framework, described in section 2, only the boundary fractal dimension was employed. The effect of the root-mean-square of the roughness height was not incorporated. Further investigation is required to understand how the root-mean-square of the roughness height might impact the predictions of our theoretical linear relationship, Eq. (7).

## 6. Conclusions

In this study, we developed a first-order linear approximation of the relationship relating fracture aperture ($d_{max}$) to fracture width ($w$). Linear relationships have been previously proposed either experimentally or theoretically in the literature. However, for the first time, we invoked concepts from fractal geometry to show that $d_{max}$ should scale linearly with $w$. We reanalyzed the Hatton et al.[20] database, for which nonlinear power-law equations, with breaks in slope, have been reported, and demonstrated that a linear relationship with various numerical prefactors could accurately capture the trends in the data. Comparison with previously published flow rate experiments also showed that our linear scaling relationship, in combination with the cubic law, could accurately scale flow rate with fracture aperture for more than 170 experimental observations.

**Acknowledgment**

BG is grateful to Kansas State University for supports through faculty startup funds. EP acknowledges support from the Army Research Laboratory under Grant Number W911NF-16-1-0043 and the Tom Cronin and Helen Sestak Faculty Achievement award. The authors acknowledge constructive comments by Ian Main, University of Edinburgh, as well as fruitful clarifications on flow rate experiments by Christian Klimczak, University of Georgia. The digitized data used in this study are available upon request from the corresponding author.

**Appendix A**

In this section, we consider an elliptical fracture with a smooth boundary presented in Fig. A1. We show that there exists no simple relationship between fracture aperture and its width, unless the latter is significantly greater than the former.

Following Eq. (3), one may relate fracture perimeter to its area for a smooth-boundary object ($D_b = 1$) as follows:

$$P \propto A^{\frac{1}{2}} \tag{A1}$$

Within Euclidean geometry, Eq. (A1) always holds because the dimension of area is [$L^2$], while that of perimeter is [L].

In the literature, various formulas have been developed to determine the perimeter of an ellipse. Most, if not all, of these are approximations. The two most widely applied models were proposed by Ramanujan[61] and are in the following forms:

$$P \approx \pi[(d_{max} + w) + \frac{3(w - d_{max})^2}{10(d_{max} + w) + \sqrt{d_{max}^2 + 14 d_{max} w + w^2)}}] \tag{A2}$$

$$P \approx \pi[3(d_{max} + w) - \sqrt{(d_{max} + 3w)(3 d_{max} + w)}] \tag{A3}$$

In addition to these relatively simple relationships, there exist several other models in series form.

In contrast to the perimeter formula, the area of an ellipse is exact and can be calculated as

$$A = \frac{\pi}{4} w d_{max} \tag{A4}$$

Combining either Eq. (A2) or (A3) with Eqs. (A1) and (A4) does not yield any straightforward relationship between $d_{max}$ and $w$ for a smooth-walled elliptical fracture. This is simply because $P$ is a complex function of $d_{max}$ and $w$ (see Eqs. A2 and A3).

Accordingly, there exists no simple relationship linking $d_{max}$ to $w$. Only if $w \gg d_{max}$, both Eqs. (A2) and (A3) reduce to a linear proportionality between perimeter and width i.e., $P \propto w$. Given that $A \propto w d_{max}$, substituting $P$ and $A$ into Eq. (A1) yields

$$d_{max} \propto w \tag{A5}$$

which is the same as our Eq. (7). The difference between Eq. (A5) and Eq. (7) is that the former holds if and only if $w \gg d_{max}$, while the latter does not depend on any assumption between $w$ and $d_{max}$.

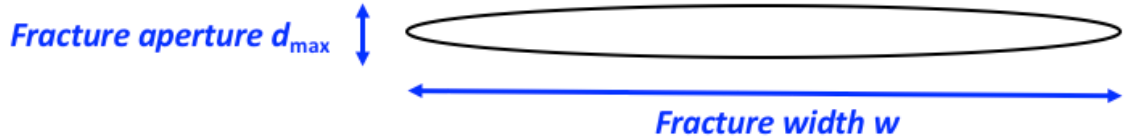

Fig. A1. An elliptical fracture with a smooth boundary. $d_{max}$ and $w$ represent the fracture aperture (maximum opening) and the fracture width, respectively.

## Appendix B

Here we assume a rectangular fracture with a fractal rough boundary (see Fig. B1) and demonstrate that under specific circumstances fracture aperture scales linearly with its width. For this purpose, following Eq. (3), we assume that

$$P \propto A^{\frac{D_b}{2}} \tag{B1}$$

If $w \gg d_{max}$, one can approximate the fracture perimeter by

$$P \propto \varepsilon^{1-D_b} w^{D_b} \tag{B2}$$

This means that the fractal rough width of the fracture dominantly contributes to the perimeter. We further assume that the area can be represented by

$$A \propto w d_{ave} \tag{B3}$$

Combining Eqs. (B2) and (B3) with Eq. (B1) gives

$$d_{ave} \propto w \tag{B4}$$

which is similar to Eq. (7). One should note that any fracture whose geometry differs significantly from that given in Fig. B1 and/or any failure in the assumptions $P \propto w^{D_b}$ (Eq. B2) and $A \propto w d_{ave}$ (Eq. B3) could cause significant deviations from Eq. (B4).

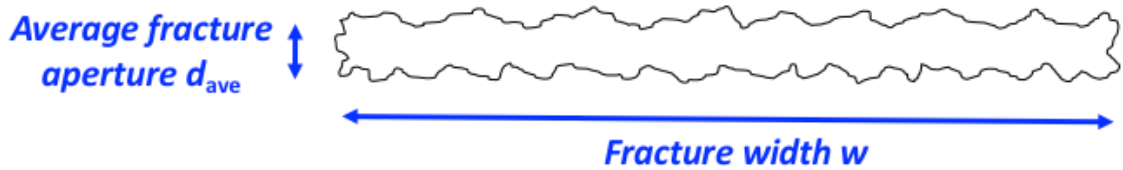

Fig. B1. A rectangular fracture with a rough boundary. $d_{ave}$ and $w$ represent the average fracture aperture and the fracture width, respectively.

**Figures caption**

Fig. 1. Schematic cross-section through an elliptical fracture with a rough boundary. $d_{max}$ and *w* represent the fracture aperture (maximum opening) and the fracture width, respectively.

Fig. 2. Fracture aperture versus fracture width for the (a) Kelduhverfi and (b) Myvatn datasets from Hatton et al.[20]. The solid lines represent Eq. (7) with various numerical prefactor values (i.e., *C* = 0.001, 0.01, and 0.1). To be consistent with Figs. (3a) and (3b) of Hatton et al.[20], same symbols have been used.

Fig. 3. Measured flow rate *Q* as a function of measured aperture for 8 datasets and 173 fractures digitized from Klimczak et al.[48]. The red and blue solid lines respectively represent Eq. (9) with *n* = 1 (from Eq. 7) and numerical prefactors $10^2$ and $10^8$ (i.e., $Q = 10^2 d_{max}^4$ and $Q = 10^8 d_{max}^4$).

Table 1. Various *n* values determined by directly fitting Eq. (2) to measurements, and their corresponding coefficients of determination ($R^2$) for several datasets reported in the literature, including theoretical and stochastic approaches.

| Reference | Remarks | No. of samples | Exponent $n^*$ | $R^2$ |
|---|---|---|---|---|
| Pollard and Segall[13] | LEFM theory | - | 1 | - |
| Scholz and Cowie[15] | Stochastic model for faults in a variety of tectonic environments | NA | 1 | NA |
| Hatton et al.[20] | Kelduhverfi, Iceland, Fracture width < 3 m | 79 | 2.2 ± 0.24 | 0.41 |
| | Kelduhverfi, Iceland, Fracture width > 3 m | | 0.89 ± 0.08 | 0.76 |
| | Myvatn, Iceland, Fracture width < 3 m | 42 | 1.78 ± 0.47 | 0.22 |
| | Myvatn, Iceland, Fracture width > 3 m | | 0.63 ± 0.08 | 0.81 |
| Vermilye and Scholz[62] | Bonticou Crag | NA | 1 | 0.96 |
| | Outcrop 1, Forillon Park, Gaspe Peninsula | NA | 1 | 0.86 |
| | Outcrop 2, Forillon Park, Gaspe Peninsula | NA | 1 | 0.94 |
| | Anse a Mercier, Gaspe Peninsula | NA | 1 | 0.73 |
| | Les Petite Anse, gaspe Peninsula | NA | 1 | 0.66 |
| | Petite Vallee, Gaspe Peninsula | NA | 1 | 0.68 |
| | White-hall Dike | NA | 1 | 0.66 |
| | Lake Champlain | NA | 1 | 0.24 |
| Walmann et al.[23] | Fractures in clay | ~3000 | 0.47 ± 0.03 | NA |
| Johnston and McCaffrey[63] | Croagh Patrick quartz veins | 431 | 1.20 | 0.73 |
| | Section parallel to the vein array, Croagh Patrick | 139 | 1.47 | 0.64 |
| | Loughshinny | 158 | 0.99 | 0.64 |
| | Loughshinny, section perpendicular to shear zone and transport direction | 180 | 1.08 | 0.32 |
| | Bridges of Ross quartz veins in sandstone | 300 | 1.30 | 0.68 |
| | Bridges of Ross, section parallel to zone | 96 | 0.95 | 0.26 |
| | Skelpoonagh Bay | 305 | 1.08 | NA |
| | Quartz veins in granite | 16 | 1.49 | 0.44 |
| Gudmundsson et al.[64] | Veins in the damage zone of the Husavik-Flatey fault | 384 | 1 | 0.66 |
| Olson[8] | Ship Rock Dike Segments from Delaney and Pollard (1981) | NA | 0.4 | 0.55 |
| | LEFM theory incorporating subcritical and critical fracture propagation criteria | - | 0.5 | - |
| Schultz et al.[46] | 14 datasets from various studies (details given in their Fig. 1) | - | 1 | NA |
| Lai et al.[65] | Open fractures in tight gas sandstones | 110 | 1 | 0.99 |
| | Calcite-filled fractures in tight gas sandstones | 300 | 1 | 0.95 |

* *n* represents the exponent in the power-law aperture-width relationship (Eq. 2); NA means not available.

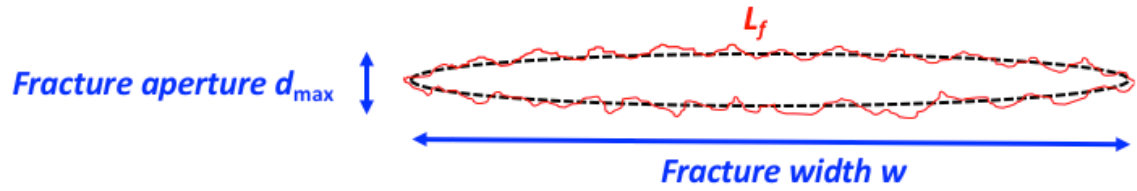

Fig. 1. Schematic cross-section through an elliptical fracture with a rough boundary. $d_{max}$ and $w$ represent the fracture aperture (maximum opening) and the fracture width, respectively.

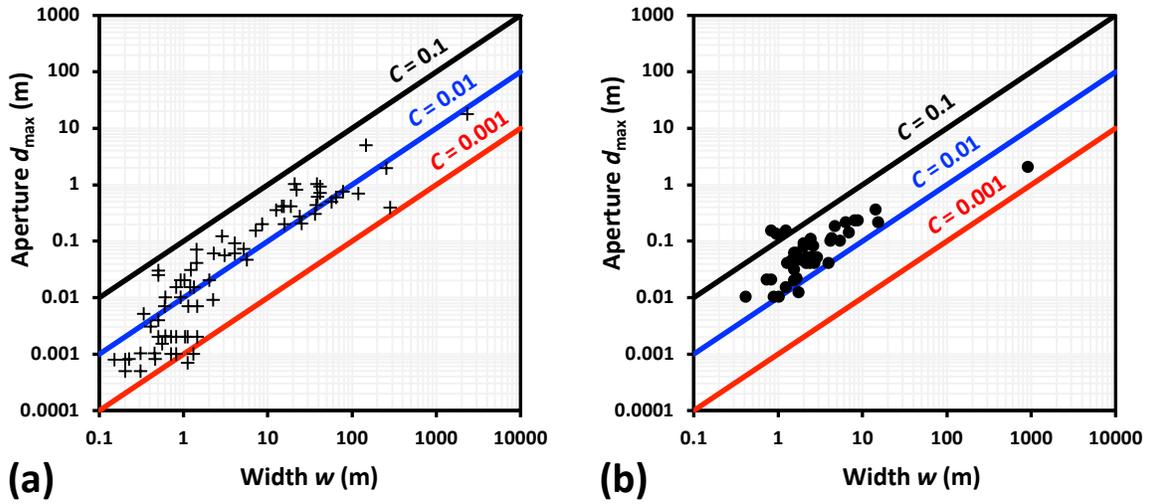

Fig. 2. Fracture aperture versus fracture width for the (a) Kelduhverfi and (b) Myvatn datasets from Hatton et al.[20]. The solid lines represent Eq. (7) with various numerical prefactor values (i.e., $C$ = 0.001, 0.01, and 0.1). To be consistent with Figs. (3a) and (3b) of Hatton et al.[20], same symbols have been used.

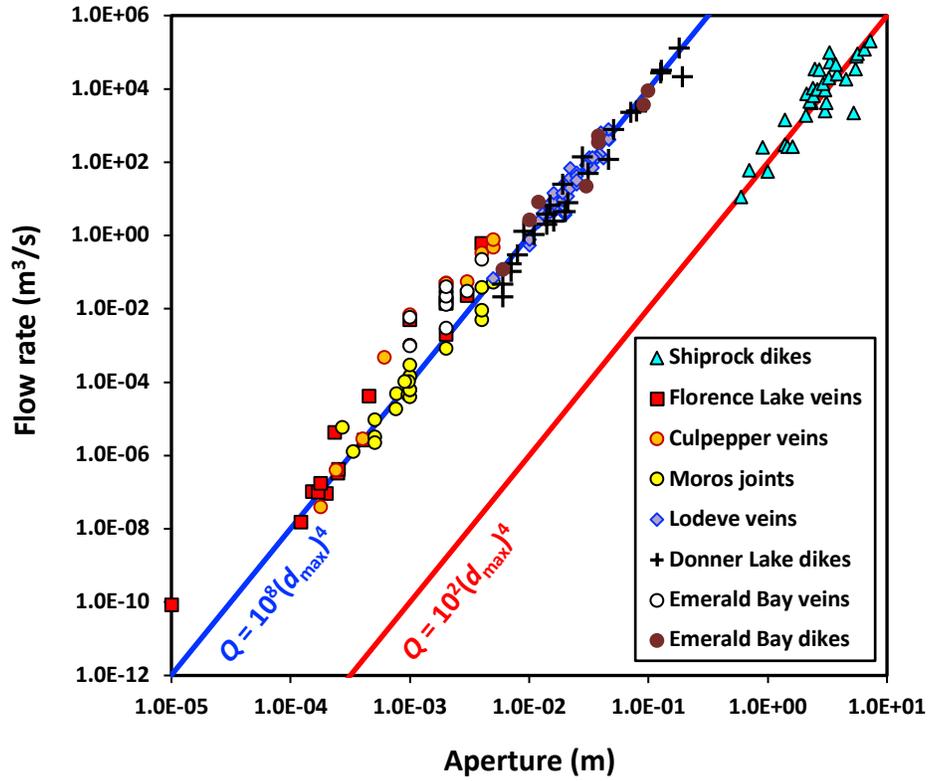

Fig. 3. Measured flow rate $Q$ as a function of measured aperture for 8 datasets and 173 fractures digitized from Klimczak et al.[48]. The red and blue solid lines respectively represent Eq. (9) with $n = 1$ (from Eq. 7) and numerical prefactors $10^2$ and $10^8$ (i.e., $Q = 10^2 d_{max}^4$ and $Q = 10^8 d_{max}^4$).